\documentclass[preprint,prd,nofootinbib,showpacs,rotate]{revtex4}

\usepackage{graphicx}
\usepackage{dcolumn}
\usepackage{bm}
\usepackage{color}
\input{epsf}
\begin{document}

\title{Static strings in Randall-Sundrum scenarios and the quark anti-quark potential}

\author{Henrique Boschi-Filho}
\email{boschi@if.ufrj.br}
\affiliation{Instituto de F\'{\i}sica, 
Universidade Federal do Rio de Janeiro, Caixa Postal 68528, 
RJ 21941-972 -- Brazil}
\author{Nelson R. F. Braga}
\email{braga@if.ufrj.br}
\affiliation{Instituto de F\'{\i}sica, 
Universidade Federal do Rio de Janeiro, Caixa Postal 68528, 
RJ 21941-972 -- Brazil}
\author{Cristine N. Ferreira}
\email{crisnfer@if.ufrj.br}
\affiliation{Instituto de F\'{\i}sica,~Universidade Federal do Rio de 
Janeiro, Caixa Postal 68528, 
RJ 21941-972 - Brazil}
\affiliation{ N\'ucleo de F\'isica, Centro Federal de Educa\c c\~ao Tecnol\'ogica de Campos,
Campos dos Goytacazes, RJ 28030-130,  Brazil}

\begin{abstract}  
We calculate the energy of a static string in an AdS slice 
between two D3-branes with orbifold condition. 
The energy for configurations with endpoints
on a brane grows linearly for large separation between these points. 
The derivative of the energy has a discontinuity at some critical 
separation.  Choosing a particular position for one of the branes 
we find configurations with smooth energy. 
In the limit where the other brane goes to infinity the energy 
has a Coulombian behaviour for short separations 
and can be identified with the Cornell potential for a quark anti-quark pair.
This identification leads to effective  
values for the AdS radius, the string tension and the position of 
the infrared brane. 
These results suggest an approximate duality 
between static strings in an AdS slice and a heavy quark anti-quark 
configuration in a confining gauge theory.

\end{abstract}


\maketitle

\vfill\eject

Phenomenological models with extra dimensions have attracted much 
attention in last years. In particular, the Randall-Sundrum model
\cite{Randall:1999ee} consists of   
a five dimensional anti-de Sitter (AdS) slice between 
two D3-branes with the standard model fields living 
on one of these branes.
Here we calculate the world-sheet area of a static string in such a space.
We find a potential energy with Coulomb like behaviour at short distances 
and linear confining behaviour for large distances.
This energy can be identified with the Cornell potential for the strong interaction
of a heavy quark anti-quark pair. 

Understanding non perturbative aspects of strong interactions,  like confinement 
and mass generation is a challenge for theoretical physicists.
At high energies, QCD (Yang Mills SU(3) plus fermionic matter fields) has a small 
coupling constant and leads to a nice perturbative description of strong interactions.  
However at low energies the coupling is large, so the theory is non perturbative
and one needs lattice calculations.

There are presently many indications that one can learn about this non perturbative regime 
of QCD from gauge/string dualities.  A very important result relating SU(N) 
Yang Mills gauge theories with large N to string theory 
was obtained by  't Hooft\cite{'tHooft:1973jz} long ago.
More recently Maldacena \cite{Maldacena:1997re} discovered an exact  gauge/string  
duality. The string theory is defined in a ten dimensional 
space which is the direct product of a five dimensional Anti-de Sitter space and a 
compact five dimensional space (AdS$_5\times$X$^5$).
The corresponding gauge theory is a superconformal large N Yang Mills  
theory on the four dimensional boundary of this space. 
This duality is known as AdS/CFT correspondence\cite{Maldacena:1997re,Gubser:1998bc,
Witten:1998qj,Aharony:1999ti}. 
A first connection between AdS/CFT and non conformal gauge theories 
was proposed by Witten in \cite{Witten:1998zw}.
In this approach the AdS space accommodates a  Schwarzschild black hole.
This procedure introduces a scale, breaking conformal invariance, and can be used,
for instance,  to calculate glueball masses\cite{MASSG, MASSG2,MASSG3, MASSG4,MASSG5,MASSG6,MASSG7}. 

Important results on strong interactions have been obtained recently
from phenomenological models inspired in the idea of gauge/string duality.
Introducing an infrared cut off in the AdS space, Polchinski and Strassler 
obtained the high energy amplitude for glueball scattering at fixed angles from 
string theory\cite{Polchinski:2001tt}. They obtained also 
results for the structure functions of deep inelastic scattering using 
this framework \cite{Polchinski:2002jw}.
This infrared cut off AdS space can be interpreted as an AdS slice, 
similar to that proposed in \cite{Randall:1999ee}. 
Using an AdS slice with boundary conditions  
the spectrum of glueballs \cite{Boschi-Filho:2002vd,Boschi-Filho:2005yh}
and light baryons and mesons \cite{deTeramond:2005su} was also obtained. 
 
In a gauge theory, non perturbative aspects such as confinement 
can be studied with the help of Wilson loops 
$\,exp \{ i \oint_C A^\mu dx_\mu \}\,$.
For a static configuration it takes the form $ exp \{ - T \,E  \, \}\,$
where $E $ is the energy and $T$ the time interval. 
Wilson loops can be used to calculate the potential energy of a heavy quark anti-quark 
pair and determine the behaviour with respect to the quark separation.
 
Gauge string/duality can be used to calculate Wilson loops.
In the case of the AdS/CFT correspondence there is an exact duality
and the Wilson loop of a heavy quark anti-quark pair in 
the superconformal large N gauge theory was calculated
from the world-sheet area of a dual static string living in 
the AdS bulk\cite{RY,MaldaPRL}. 
The string lies along a geodesic in the AdS bulk
with endpoints on the boundary representing the quark and anti-quark positions.
Its energy is proportional to the geodesic length
and is a function of the quark anti-quark  distance 
for an observer on the four dimensional boundary where the gauge theory is defined. 
In refs. \cite{RY,MaldaPRL} it was shown that for the AdS$_5$ space  
the energy shows a purely Coulombian (non confining) behaviour, compatible 
with a conformal field theory. 
For calculations of Wilson loops following Witten's proposal\cite{Witten:1998zw} for a confining geometry see for instance\cite{Greensite:1998bp, Greensite:1999jw,Bigazzi:2004ze,Martucci:2005yg}.
A discussion of Wilson loops associated with quark anti-quark potential
in general spaces can be found in \cite{Kinar:1998vq}. 

Here we are going to calculate the potential energy of a static string in an AdS slice 
with orbifold condition as  in the Randall Sundrum model. 
In this model the standard model fields live on a 
$D3$-brane ($ y=0$) in a five dimensional space described by the metric
\begin{equation}
\label{1.1}
ds^2 = e^{-2 \vert y  \vert / R} \eta_{\mu \nu} dx^{\mu} dx^{\nu } +  dy^2
\end{equation}

\noindent
where $\eta_{\mu \nu}$ is the four dimensional 
Minkoviski metric. The coordinate $y $ is defined in the range 
 $ - y_c \le y \le  y_c \, $, where $y_c\,$ is a constant related to the 
compactification length and satisfies an orbifold condition corresponding 
to the identification: $ ( x , y )\,=\,( x , - y ) $. 
This space corresponds to two identical AdS slices joined together with endpoints 
of the coordinate $y$ identified.  
Note that we just need to consider one slice since the other is just a copy.
It is convenient to describe the slice $ y \ge 0 $  using the coordinates 
$ x , r\,$ with $ r \,=\, R \exp \{-  y   / R\} $ which implies
\begin{equation}
\label{metric}
ds^2 \,=\, \Big( {r^2\over R^2} \Big) ( -dt^2 + d{\vec x}^2 ) +  
\Big( {R^2\over r^2} \Big) dr^2 \,.
\end{equation}

\noindent with 
$ r_{2}\,=\,R \exp \{-  y_c  / R\} \le r \le r_1 = R $.

We want to calculate the potential energy associated with a static 
string with endpoints located on the standard model brane ($r_1 = R$). 
The string is described by the Nambu-Goto action  
\begin{equation}
S \,=\,{1\over 2\pi \alpha^\prime}\, \int d\sigma d\tau 
\sqrt{ det \big( g_{MN} \partial_\alpha X^M 
\partial_\beta X^N \Big)\,\,}\,
\end{equation}

\noindent where $g_{MN}$ is the five dimensional metric defined by (\ref{metric}). 
This action is proportional to the area of the world-sheet. For a static 
configuration the action is also proportional to the energy. 
Solving the classical equations of motion, that corresponds to finding
the geodesics, one can estimate the energy from the string length. 

Let us consider the geodesic between two points on the brane at $r_1 = R$ 
separated by a coordinate distance $\Delta x\,=\, L$. 
Since there is a brane at $r = r_{2} $, there will be
two classes of geodesics solutions, as illustrated in fig. 1. 
For coordinate separations smaller than some critical value $ L_{crit}\,$ 
the geodesic does not reach $r_{2} $. 
In this case the geodesics reaches a minimum value $r_0$ of
the coordinate $r$ defined by the relation
\begin{equation}
L \,=\,\frac{ 2 R^2 }{r_0}\,I_1 (R/r_0)
\end{equation}

\noindent where $I_1 (\xi )$ is the elliptic integral
\begin{equation}
I_1 (\xi )\,=\,\int_1^{\,\xi}\,
\frac{ d \rho }{\rho^2 \,\sqrt{ \rho^4 -1}}\,.
\end{equation}

\noindent The critical coordinate separation $L\,=\,L_{crit}\,$ corresponds to 
$r_0\,=\,r_{2}\,$. For $L > L_{crit} $ the geodesic reaches the second brane 
and contains a non null path along  $r\,=\,r_{2} $. 
This happens because this region corresponds to a minimum of the metric thanks to 
the orbifold condition.

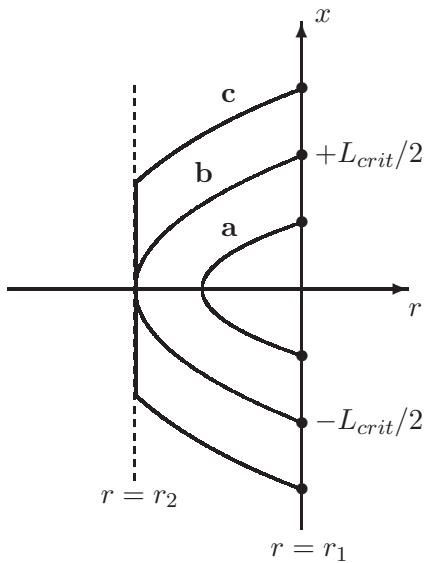
\begin{figure}
\centering

\
\setlength{\unitlength}{0.07in}
\vskip 5.5cm
{\begin{picture}(0,0)(18,0)
\rm\thicklines\bf
\put(25,-8){\vector(0,0){38}}
\put(26,-0.5){$- L_{crit}/2$}
\put(26,19.5){$+L_{crit}/2$}
\put(26,30){$x$}
\put(24.5,-0.5){$\bullet$}
\put(24.5,19.5){$\bullet$}
\put(24.5,14.5){$\bullet$}
\put(24.5,4.5){$\bullet$}
\put(24.5,24.5){$\bullet$}
\put(24.5,-5.5){$\bullet$}
\put(17,18){b}
\put(19,14){a}
\put(19,24){c}
\put(3,10){\vector(1,0){30}}
\put(33,8){$r $}
\put(22.7,-10 ){$r = r_1 $}
\put(10,-6){$r = r_2$}
\bezier{600}(25,15)(10,10)(25,5)
\bezier{600}(25,20)(0,10)(25,0)
\bezier{600}(25,25)(17,22)(12.6,18)
\bezier{600}(25,-5)(17,-2)(12.6,2)
\bezier{600}(12.5,18)(12.5,10)(12.5,2)
\bezier{600}(12.55,18)(12.55,10)(12.55,2)
\multiput(12.55,-4.3)(0,1){30}{\line(0,1){0.5}}
\end{picture}}
\vskip 2.5cm
\vskip .5cm
 \parbox{5in}{\caption{ Schematic representation 
of geodesics in the Randall Sundrum space. Curve {\bf a } 
corresponds to a geodesic with $L < L_{crit}$, curve {\bf b} to 
$L = L_{crit}$ and {\bf c} to $L > L_{crit}$.}}
\end{figure}

The string energy is proportional to the geodesic length.
For $L \le L_{crit}\,$ it can be calculated following 
refs. \cite{Kinar:1998vq,Boschi-Filho:2004ci} obtaining 
\begin{equation}
E_{RS}^{\,(-)} \,=\, \frac{ r_0 }{\pi \alpha^\prime} I_2 ( R/r_0 ) 
\,-\,\frac{ r_0 }{\pi \alpha^\prime} 
\end{equation}

\noindent where we have subtracted the constant $ R /\pi \alpha^\prime\,$ 
for latter convenience and defined a second elliptic integral
\begin{equation}
I_2 (\xi) \,=\, \int_1^{\xi}\,
\Big[\,\frac{ \rho^2 }{\sqrt{ \rho^4 -1}} \,-\,1 \,\Big] d\rho \,.
\end{equation}
 
The energy corresponding to the geodesics for $L \ge L_{crit}\,$ is
calculated by adding the lengths of curves between $r =r_2$ and 
$r = R$ and the path along the brane at $r = r_2$ (see figure 1, curve {\bf c}). 
After subtracting the same constant we obtain
\begin{equation}
E_{RS}^{\,(+)} \,=\, \frac{ r_0 }{\pi \alpha^\prime} 
\Big( I_2 (R/r_0) - I_2 (r_2/r_0)\,\Big)
\,-\,\frac{ r_2 }{\pi \alpha^\prime} \,+\,
 \frac{ r_2^2 }{\pi \alpha^\prime\,r_0} 
I_1 ( r_2/r_0). 
\end{equation}

We note that for large values of $L$ the energy increases linearly with $L$
(confining behaviour).
However, the energy is not a smooth function of $L$ in this model. 
There is a discontinuity in the derivative at $L = L_{crit}$.
See figure 2. 
This discontinuity decreases as we increase the value of $r_2$.
If we take $r_2 \to R $ the discontinuity disappears.
However this corresponds to placing the two branes at the same position
and then having no AdS slice.

\begin{figure}
\centering
\includegraphics[width=8cm]{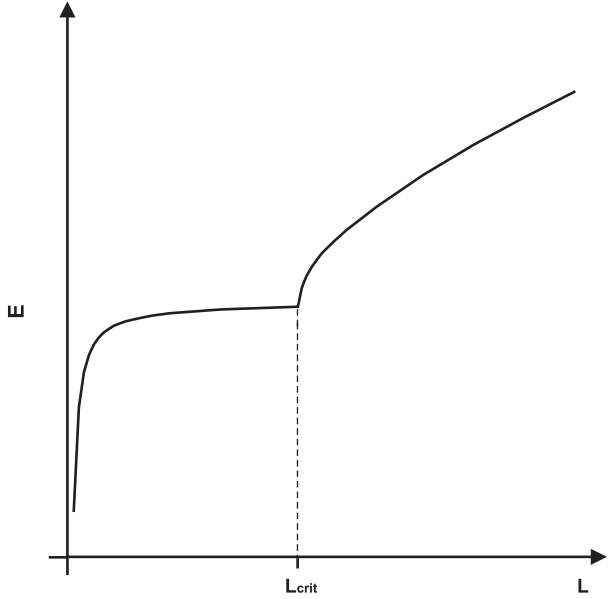}
\parbox{5in}{\caption{ Energy as a function of string end-points separation $L$ 
in Randall Sundrum space $r_2 \le r \le R $ with discontinuous derivative at $L_{crit}$.} }
\end{figure}

We can modify our initial set up to find an energy that varies smoothly 
with the endpoints separation. 
We consider the standard model brane at some $r_1 > R $ and the second brane 
at some $r_2 $. We find out that the energy is smooth only if $r_2 = R$. 
So we take two AdS slices with orbifold identification
each one described by the metric (\ref{metric}) but now with 
$ r_2 = R \le r \le r_1 $. In this case the relation between $L$ and $r_0$ is
\begin{equation}
\label{Lr1}
L \,=\,\frac{ 2 R^2 }{r_0}\,I_1 (r_1/r_0 )
\end{equation}
 
\noindent The energy for $L \le L_{crit}\,$ can be defined subtracting the 
constant $ r_1 /\pi \alpha^\prime\,$ in such a way that the energy is finite 
even in the limit $r_1 \to \infty $. We get  
\begin{equation}
E^{\,(-)} \,=\, \frac{ r_0 }{\pi \alpha^\prime} \,
\Big[ \, I_2 (r_1/r_0)\,-\,1 \Big] 
\end{equation}

\noindent Substituting $r_0$ from eq. (\ref{Lr1}) we find
\begin{equation}
\label{E-}
E^{\,(-)} \,=\, \frac{ 2 R^2 }{\pi \alpha^\prime} \,\frac{I_1 (r_1/r_0 )}{L}
\Big[ \, I_2 (r_1/r_0)\,-\,1 \Big] \,.
\end{equation}

For $ L > L_{crit} $, again subtracting the constant $ r_1 /\pi \alpha^\prime\,,$ 
we obtain
\begin{eqnarray}
\label{Er1}
E^{\,(+)} &=& \frac{ r_0 }{\pi \alpha^\prime} 
\Big[ I_2( r_1/r_0) - I_2 (R/r_0)\Big]  
\,-\,\frac{ R }{\pi \alpha^\prime} \,+\,
 \frac{ R^2 }{\pi \alpha^\prime\,r_0} 
I_1 ( R/r_0)  \\
\nonumber\\
\label{Er12}   
&=& \frac{ 2 R^2 }{ \,\pi \alpha^\prime} \frac {I_1 ( r_1/r_0)}{L}
\Big[ I_2( r_1/r_0) - I_2 (R/r_0)\Big]
\,-\,\frac{ R }{\pi \alpha^\prime} \,+\,
 \frac{ L  }{2 \pi \alpha^\prime\,} 
\frac {I_1 ( R/r_0)}{ I_1 (r_1/r_0)} \,,
\end{eqnarray}

\noindent where again we have substituted $r_0$ from  eq. (\ref{Lr1}).

It is interesting to analyze the behaviour of these energies 
in the limit where the standard model brane is moved to infinity 
$\,r_1 \to \infty \,$.
Noting that  $ I_1 (\infty )\,\equiv\,C_1\,=
\,\sqrt{2}\pi^{3/2}/[\Gamma(1/4)]^2\,\,$ 
and $\,\,I_2 (\infty ) \,=\,1\,-\,C_1 $,
we find that the energy (\ref{E-}) 
behaves as $\sim 1/L$ and the energy (\ref{Er1}) 
shows a linear behaviour for large values of $L$.   

These asymptotic behaviors are analogous to those of the
phenomenological Cornell potential for a heavy quark anti-quark pair 
\cite{Quigg:1977dd,Eichten:1978tg,Martin:1980jx}
\begin{equation}
\label{Cornell}
E_{Cornell}(L) \,=\, -\frac{4}{3} \frac{a}{L} \,+\, \sigma L\,+\, const.\,\,.
\end{equation}

\noindent where $L$ is the quark separation, $ a = 0.39 $ and $\sigma\,=\, 0.182$ Gev$^2$. For a review see for instance 
\cite{Brambilla:1999ja,Nesterenko:1999dx}. 

This kind of potential, with a linear and a
Coulombian terms, has been also obtained from Wilson loops in other confining backgrounds \cite{Greensite:1999jw,Bigazzi:2004ze,Martucci:2005yg}.  

The fact that our potential behaves asymptotically as the Cornell potential 
suggests that there is an approximate gauge/string duality 
relating strings in this AdS slice with a QCD like confining gauge theory.
In particular, the world-sheet area of the static string considered here would be dual
to the Wilson loop of a heavy quark anti-quark pair in the dual gauge theory. 
So the string endpoints correspond to the quark and anti-quark positions
and $L$ is the quark anti-quark distance from the point of view of the gauge theory.

Then we can identify the asymptotic behaviour of the string energy $E^{\,(-)}$
for small $L$ with the Coulomb term of the Cornell potential by taking 
$ a\,=\,3 C_1^2 R^2/2\pi \alpha'$.
In the same way we can identify the asymptotic behaviour of
$E^{\,(+)}$ for large $L$ with the linear term of the Cornell potential 
by choosing  $\sigma = 1/2\pi\alpha'$.

So the string energy as a function of end-point separation 
$L$ in terms of the Cornell parameters is
\begin{eqnarray} 
E &=&
\left\{ \matrix { \displaystyle \frac{ 4 a }{ 3 C_1^2} \,\frac{I_1 (r_1/r_0 )}{L}
\Big[ \, I_2 (r_1/r_0)\,-\,1 \Big]\,, \hskip 5cm \qquad L \le L_{crit} \cr \cr 
\displaystyle \frac{ 4 a }{ 3 C_1^2} \frac {I_1 ( r_1/r_0)}{L}
\Big[ I_2( r_1/r_0) - I_2 (R/r_0)\Big]
\,-\sqrt{\frac{\, 4 \,a \,\sigma }{ 3\, C_1^2}} \,+\,
 \sigma  L \,\,\frac {I_1 ( R/r_0)}{ I_1 (r_1/r_0)}\,, \qquad  L \ge L_{crit}
} \right.\nonumber\\
\end{eqnarray}
   
In the limit $r_1 >> R$, where according to eq. (\ref{Lr1}), $L_{crit} \to 2 R C_1 $ the energy takes the form
\begin{equation} 
E \,=\,
\left\{ \matrix { \displaystyle \,-\, \frac{ 4 a }{ 3 \,L } \,
\,, \hskip 8.5cm \qquad L \le L_{crit} \cr \cr 
\displaystyle - \frac{ 4 a }{ 3 L}
\,+\,\frac{ 4 a }{ 3 C_1 \,L} 
\Big[ 1 - I_2 (R/r_0)\Big]
\,-\sqrt{\frac{\, 4 \,a \,\sigma }{ 3\, C_1^2}} \,+\,
 \sigma  L \,\,\frac {I_1 ( R/r_0)}{ C_1 }\,, \qquad L \ge L_{crit}
} \right.
\end{equation}
   
\bigskip

\noindent Note that for $L \le L_{crit}\,$ the string does not reach the infrared brane.
So, the potential takes the Coulombian form corresponding to a conformal theory
as in AdS/CFT case.
For $L >> R$ the energy takes the asymptotic
linear form $ E \sim \sigma \,L\,$ as in the Cornell potential.
From the point of view of the gauge theory this implies confinement.
In figure 3 we represent the energy for different brane 
positions $r_1 = nR$ and $r_2 = R$ fixed.

\begin{figure}
\centering
\includegraphics[width=8cm]{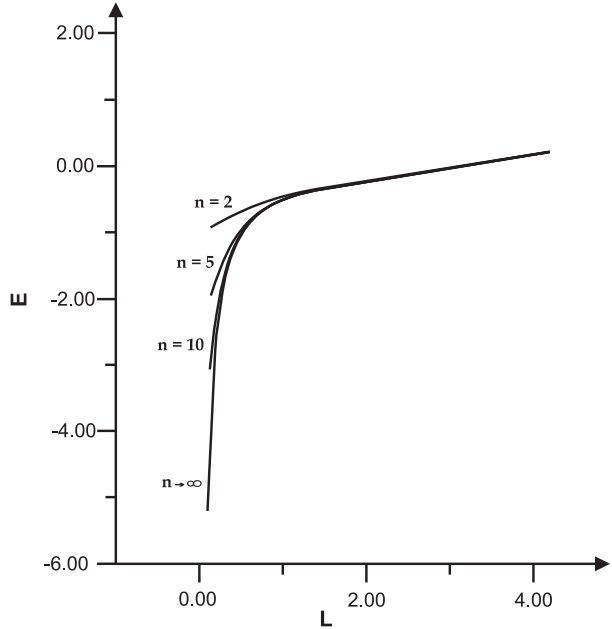}
\parbox{5in}{\caption{ Energy in GeV as a function of string end-points 
separation $L$ in GeV$^{-1}$ for AdS slices with
 $r_1 = nR $ and $r_2 = R $ . For $ n \to \infty$  
the energy behaves as the Cornell potential eq. (\ref{Cornell}).}}
\end{figure}

The identifications we have done for the energy in the Randall Sundrum scenarios 
(an AdS slice between two branes $ r_2=R \le r \le r_1 $) determines an effective value 
for the AdS radius in this phenomenological model in terms of the Cornell parameters:
\begin{equation}
R \,=\,\sqrt{ \frac{a}{ 3 \sigma C_1^2}}\,=\, 1.4\,\, {\rm GeV}^{-1}\,\,.
\end{equation}

\noindent This corresponds to an effective energy scale of $0.71$ GeV 
consistent with a QCD scale.

Concluding, we calculated static string energies in AdS slices with orbifold
condition and found smooth energies when the infrared brane is located at $r_2 = R$.
In the limit where the standard model brane goes to infinity 
the potential energy is identified with the Cornell potential for a heavy
quark anti-quark pair. This fact suggests a duality between string theory 
in this AdS slice and a confining gauge theory on the boundary.
The string world-sheet area would be dual to the static gauge theory Wilson loop. 
The identification of the string energy with the Cornell potential
fixes an effective AdS radius, the position of the infrared brane 
and the string tension.

Recent results in the literature also support the idea of this approximate duality.
An AdS slice, or equivalently an AdS space with an infrared cut off, was used in refs. 
\cite{Polchinski:2001tt,Polchinski:2002jw,GI,BB3,BT,AN,Brodsky:2003px}
to discuss hadronic scattering from a string point of view.
It has also been used to obtain Regge trajectories and hadron masses
 \cite{Boschi-Filho:2002vd,deTeramond:2005su,Boschi-Filho:2005yh} and 
masses, decay rates and couplings for the lightest mesons\cite{Erlich:2005qh}.

\newpage
     
\noindent {\bf Acknowledgments}: We would like to thank Erasmo Ferreira
for important discussions. The authors are partially supported by CNPq and Faperj.


\begin{thebibliography}{30}
 
\bibitem{Randall:1999ee}
  L.~Randall and R.~Sundrum,
  Phys.\ Rev.\ Lett.\  {\bf 83}, 3370 (1999);
ibid.  {\bf 83}, 4690 (1999).

\bibitem{'tHooft:1973jz}
  G.~'t Hooft,
  Nucl.\ Phys.\ B {\bf 72}, 461 (1974).

\bibitem{Maldacena:1997re}
  J.~M.~Maldacena,
  Adv.\ Theor.\ Math.\ Phys.\  {\bf 2}, 231 (1998).

\bibitem{Gubser:1998bc}
  S.~S.~Gubser, I.~R.~Klebanov and A.~M.~Polyakov,
  Phys.\ Lett.\ B {\bf 428}, 105 (1998).

\bibitem{Witten:1998qj}
  E.~Witten,
  Adv.\ Theor.\ Math.\ Phys.\  {\bf 2}, 253 (1998).

\bibitem{Aharony:1999ti}
  O.~Aharony, S.~S.~Gubser, J.~M.~Maldacena, H.~Ooguri and Y.~Oz,
  Phys.\ Rept.\  {\bf 323}, 183 (2000).

\bibitem{Witten:1998zw}
  E.~Witten,
  Adv.\ Theor.\ Math.\ Phys.\  {\bf 2}, 505 (1998).

\bibitem{MASSG} C. Csaki, H. Ooguri, Y. Oz and J. Terning, JHEP {\bf 9901}, 017 (1999).

\bibitem{MASSG2} R. de Mello Koch, A. Jevicki, M. Mihailescu , J. P. Nunes,  
Phys.Rev. {\bf D58}, 105009 (1998).

\bibitem{MASSG3} A. Hashimoto , Y. Oz , Nucl.Phys. {\bf B548}, 167 (1999). 

\bibitem{MASSG4} C. Csaki , Y. Oz , J. Russo , J. Terning , 
Phys.Rev. {\bf D59}, 065012 (1999). 

\bibitem{MASSG5} J. A. Minahan,  JHEP {\bf 9901}, 020 (1999).

\bibitem{MASSG6} C. Csaki, J. Terning,  AIP Conf. Proc. {\bf 494}, 321 (1999).
  
\bibitem{MASSG7} R. C. Brower, S. D. Mathur , C. I. Tan , Nucl.Phys.B587, 249 (2000). 

\bibitem{Polchinski:2001tt}
  J.~Polchinski and M.~J.~Strassler,
  Phys.\ Rev.\ Lett.\  {\bf 88}, 031601 (2002).

\bibitem{Polchinski:2002jw}
  J.~Polchinski and M.~J.~Strassler,
 JHEP {\bf 0305}, 012 (2003).

\bibitem{Boschi-Filho:2002vd}
H.~Boschi-Filho and N.~R.~F.~Braga,
 JHEP {\bf 0305}, 009 (2003);
 Eur.\ Phys.\ J.\ C {\bf 32}, 529 (2004).

\bibitem{Boschi-Filho:2005yh}
 H.~Boschi-Filho, N.~R.~F.~Braga and H.~L.~Carrion,
Phys.\ Rev.\ D {\bf 73}, 047901 (2006).

\bibitem{deTeramond:2005su}   G.~F.~de Teramond and S.~J.~Brodsky,
Phys. Rev. Lett. 94, 201601 (2005).

\bibitem{RY} S.~J.~Rey and J.~T.~Yee,
Eur.\ Phys.\ J.\ C {\bf 22}, 379 (2001).
  
\bibitem{MaldaPRL} J.~Maldacena, 
Phys.\ Rev.\ Lett.\  {\bf 80}, 4859 (1998).

\bibitem{Greensite:1998bp}
  J.~Greensite and P.~Olesen,
  JHEP {\bf 9808}, 009 (1998).

\bibitem{Greensite:1999jw}
  J.~Greensite and P.~Olesen,
  JHEP {\bf 9904}, 001 (1999).

\bibitem{Bigazzi:2004ze}
  F.~Bigazzi, A.~L.~Cotrone, L.~Martucci and L.~A.~Pando Zayas,
  Phys.\ Rev.\ D {\bf 71}, 066002 (2005).

\bibitem{Martucci:2005yg}
  L.~Martucci,
  Fortsch.\ Phys.\  {\bf 53}, 936 (2005).

\bibitem{Kinar:1998vq}
Y.~Kinar, E.~Schreiber and J.~Sonnenschein,
Nucl.\ Phys.\ B {\bf 566}, 103 (2000).

\bibitem{Boschi-Filho:2004ci}
  H.~Boschi-Filho and N.~R.~F.~Braga,
  JHEP {\bf 0503}, 051 (2005).

\bibitem{Quigg:1977dd}
  C.~Quigg and J.~L.~Rosner,
  Phys.\ Lett.\ B {\bf 71}, 153 (1977).

\bibitem{Eichten:1978tg}
  E.~Eichten, K.~Gottfried, T.~Kinoshita, K.~D.~Lane and T.~M.~Yan,
Phys.\ Rev.\ D {\bf 17}, 3090 (1978).
[Erratum-ibid.\ D {\bf 21}, 313 (1980)].

\bibitem{Martin:1980jx}
  A.~Martin,
 Phys.\ Lett.\ B {\bf 93}, 338 (1980).

\bibitem{Brambilla:1999ja}
  N.~Brambilla and A.~Vairo,
  arXiv:hep-ph/9904330.

\bibitem{Nesterenko:1999dx}   A.~V.~Nesterenko,
 Phys.\ Rev.\ D {\bf 62}, 094028 (2000).

\bibitem{GI} 
S. B. Giddings, 
Phys.\ Rev.\ D 67, 126001 (2003). 

\bibitem{BB3} H. Boschi-Filho and N. R. F. Braga, Phys. Lett. B560, 232 (2003).

\bibitem{BT} 
R. C. Brower , C-I Tan, 
Nucl.\ Phys.\ B 662, 393 (2003).

\bibitem{AN}   O.~Andreev,
  Phys.\ Rev.\ D {\bf 67}, 046001 (2003).

\bibitem{Brodsky:2003px}
  S.~J.~Brodsky and G.~F.~de Teramond,
  Phys.\ Lett.\ B {\bf 582}, 211 (2004).

\bibitem{Erlich:2005qh}
  J.~Erlich, E.~Katz, D.~T.~Son and M.~A.~Stephanov,
Phys.Rev.Lett. 95, 261602 (2005). 

\end{thebibliography}
\end{document}